# Cavity-enhanced Raman microscopy of individual carbon nanotubes

Thomas Hümmer[1,2], Jonathan Noe[1,3], Matthias S. Hofmann[1,3], Theodor W. Hänsch[1,2], Alexander Högele[1,3] & David Hunger[1,2]

Raman spectroscopy reveals chemically specific information and provides label-free insight into the molecular world. However, the signals are intrinsically weak and call for enhancement techniques. Here, we demonstrate Purcell enhancement of Raman scattering in a tunable high-finesse microcavity, and utilize it for molecular diagnostics by combined Raman and absorption imaging. Studying individual single-wall carbon nanotubes, we identify crucial structural parameters such as nanotube radius, electronic structure and extinction cross-section. We observe a 320-times enhanced Raman scattering spectral density and an effective Purcell factor of 6.2, together with a collection efficiency of 60%. Potential for significantly higher enhancement, quantitative signals, inherent spectral filtering and absence of intrinsic background in cavity-vacuum stimulated Raman scattering render the technique a promising tool for molecular imaging. Furthermore, cavity-enhanced Raman transitions involving localized excitons could potentially be used for gaining quantum control over nanomechanical motion and open a route for molecular cavity optomechanics.

[1] Fakultät für Physik, Ludwig-Maximilians-Universität, Schellingstraße 4, München 80799, Germany. [2] Max-Planck-Institut für Quantenoptik, Hans-Kopfermann-Straße 1, Garching 85748, Germany. [3] Center for NanoScience (CeNS), Ludwig-Maximilians-Universität, Schellingstraße 4, München 80799, Germany. Correspondence and requests for materials should be addressed to A.H. (email: alexander.hoegele@lmu.de) or to D.H. (email: david.hunger@physik.lmu.de).

Raman spectroscopy is a powerful technique to identify the molecular structure of a sample. Combined with microscopy, it can provide high-resolution images with chemical contrast. This offers vast potential to investigate a broad variety of micro and nanosystems of physical, chemical and biological relevance. However, the inherently weak scattering cross-section of Raman transitions poses a challenge, and is particularly severe for spatially resolved measurements of single- or few-particle levels. Therefore, various enhancement techniques have been devised, which either rely on stimulating radiation[1] or near-field enhancement[2,3].

Purcell enhancement in optical microcavities[4] was suggested early[5,6] for its use in Raman spectroscopy. It relies on the stimulating effect of increased vacuum fluctuations within a microcavity to enhance the scattering process. It thus does not require stimulating radiation, and thereby avoids high-intensity levels at the sample, and does not introduce any intrinsic background. Detailled knowledge of the cavity mode structure allows to quantitatively evaluate the observed signal strength, which is difficult, for example, for near-field enhancement techniques. However, due to the need for microscopic mode volumes, Purcell enhancement of Raman signals has only been observed in monolithic systems with limited spatial control or spectral tunability like microdroplets[5,7], photonic crystal cavities[8] and Fabry–Pérot microcavities[6,9–13]. This has restricted the spectral coverage and impeded spatially resolved measurements so far, and only large ensembles of molecules have been studied.

Here, we present Purcell-enhanced Raman imaging using an open-access fibre-based microcavity[14]. This cavity platform provides small mode volume and large quality factor leading to significant Purcell enhancement, and enables spectrally and spatially resolved measurements. We demonstrate cavity-enhanced Raman imaging to study individual single-walled carbon nanotubes (CNTs), a material with widespread applications in fields as diverse as electronics[15], photonics[16], nanomechanics[17] and quantum optics[18,19]. To obtain information on a CNT sample's heterogeneity with regards to their chirality, crystalline quality and local environment one has to characterize multiple isolated CNTs that are spatially well-separated. Raman imaging is particularly sensitive to these differences[20], and combined with complementary measurements probing, for example, absorption[21–23] can enable a detailed characterization. Notably, cavity-enhancement is at the same time present for Raman scattering, absorption and fluorescence[24–26]. Our approach thus allows to simultaneously probe the vibrational spectrum and the optical properties of a wide range of samples with enhanced sensitivity.

## Results

**Scanning-cavity setup.** We use a high-finesse fibre-based Fabry–Pérot microcavity[14], where the laser-machined concave end-facet of an optical fibre is used as a micro-mirror, see Fig. 1a. Altogether with a planar macroscopic mirror that carries the sample and that can be moved laterally, we realize a scanning-cavity microscope[26–28]. To record spatially resolved measurements, we raster-scan the large mirror and detect cavity signals such as the resonant cavity transmission or Purcell-enhanced Raman scattering pixel-by-pixel. The fibre mirror is mounted on a piezo actuator that allows to continuously set the cavity length to achieve suitable resonance conditions (see the 'Methods' section). The dielectric mirror coatings lead to a finesse of up to $\mathcal{F} = 56,000$ and a quality factor $Q_c = 2\mathcal{F}d/\lambda$ depending on the cavity length $d$. A small radius of curvature of the concave micro-mirror on the fibre and a few-micron cavity length enables small mode volumes down to $V = 13\lambda^3$ (see the 'Methods' section).

**Purcell enhancement of Raman scattering.** Cavity-enhanced Raman scattering can be understood analogously to fluorescence emission in a resonant cavity[4,29]. The spectral density of the Raman scattered light is enhanced by the (ideal) Purcell factor $C_0$ due to increased vacuum fluctuations (proportional to $V^{-1}$) and an increased density of states (proportional to $Q_c$), combining to $C_0 = (3)/(4\pi^2)\lambda^3(Q_c)/(V)$. For our cavities with $Q_c \approx 10^5$ and $V \approx 10\lambda^3$, the enhancement of the spectral density of the emission reaches values $C_0 \approx 10^3$. In the regime where the cavity resonance is more narrow than a particular Raman feature, the effective Purcell factor $C = (3)/(4\pi^2)\lambda^3(Q_{eff})/(V)$ is the figure of merit[30], which accounts for the fact that the cavity enhances only part of the transition. This is included by replacing $Q_c$ with the effective quality factor $Q_{eff} = (Q_c^{-1} + Q_r^{-1})^{-1}$ (ref. 31), where $Q_r$ is the quality factor of the Raman feature when assuming a Lorentzian spectrum. Figure 1b shows a schematic level scheme of cavity-enhanced Raman scattering and Figure 1d shows a calculation of the ideal and effective Purcell enhancement as a function of the cavity mode volume for different $Q_c$. Typical Raman features of CNTs have $Q_r \approx 1,000–2,000$ such that for a cavity with $V \approx \lambda^3$ one can expect $C \approx 100$. With narrower Raman lines one can approach $C = C_0$. The cavity-enhanced Raman field is scattered entirely into the easily collected cavity mode, which, for an asymmetric choice of mirror transmissions ($T_2 \gg T_2$) and low loss ($L \ll T_2$; see the 'Methods' section), can yield a near-unity outcoupling efficiency from the cavity $\eta_c = T_2/(T_1 + T_2 + L)$.

**Cavity-enhanced extinction microscopy.** For our experiments we transfer single-wall CNTs with an average diameter of 1.4 nm grown by chemical vapour deposition onto the planar mirror (see the 'Methods' section). The mirror is designed to have a field maximum at the surface to ensure optimal coupling of nanotubes to the cavity mode. We study two different cavities (A and B, see the 'Methods' section) with coatings for different wavelength ranges. In a first step, we perform scanning-cavity-extinction microscopy[26] on a large area of the sample. The cavity enhances the extinction signal by a factor $4\mathcal{F}/\pi = 7 \times 10^4$ (cavity A), where $\mathcal{F}$ is the cavity finesse, and thereby enables superior sensitivity for absorption microscopy. To acquire an image, we raster-scan the large mirror and tune the mirror separation over a few hundred nanometres at each pixel to sample the cavity resonance at the wavelength of maximum finesse with a narrow-band probe laser (780 nm for cavity A and 890 nm for cavity B, respectively, see the 'Methods' section). We record the resonant cavity transmission $T_c$ on a photodiode to infordate the sample extinction $B$ from $T_c = \epsilon 4T_1 T_2/(T_1 + T_2 + L_1 + L_2 + 2B)^2$. Here, $\epsilon$ is the mode matching between the fibre mode and the cavity mode and $T_{1,2}, L_{1,2}$ are the respective mirror transmission and loss, which are all determined independently (see the 'Methods' section). From the sample-induced loss we deduce the extinction cross-section $\sigma_{ext} = B\pi w_0^2/4$, where $w_0$ is the $e^{-1}$ cavity field mode waist radius. To account for the anisotropy of the CNT extinction, we take four measurements with different orientations of linear polarization and evaluate the maximum extinction cross-section observed for each pixel. Alternatively, we use circularly polarized light (see the 'Methods' section). Figure 2d shows a measurement with cavity A where we quantitatively image the extinction cross-section of individual CNTs with a background noise level of 5 nm². We observe continuous elongated structures where a dominant fraction shows an extinction cross-section per length of $\sigma_1 = \sqrt{2/\pi}\sigma_{ext}/w_0 = 45$ nm² $\mu$m$^{-1}$ or a corresponding cross-section per carbon atom of $3 \times 10^{-4}$ nm$^{-2}$ when assuming a 1.4-nm tube diameter. This is consistent with individual CNTs probed away from an absorption resonance[21,22] and demonstrates the potential for quantitative analysis of our technique.

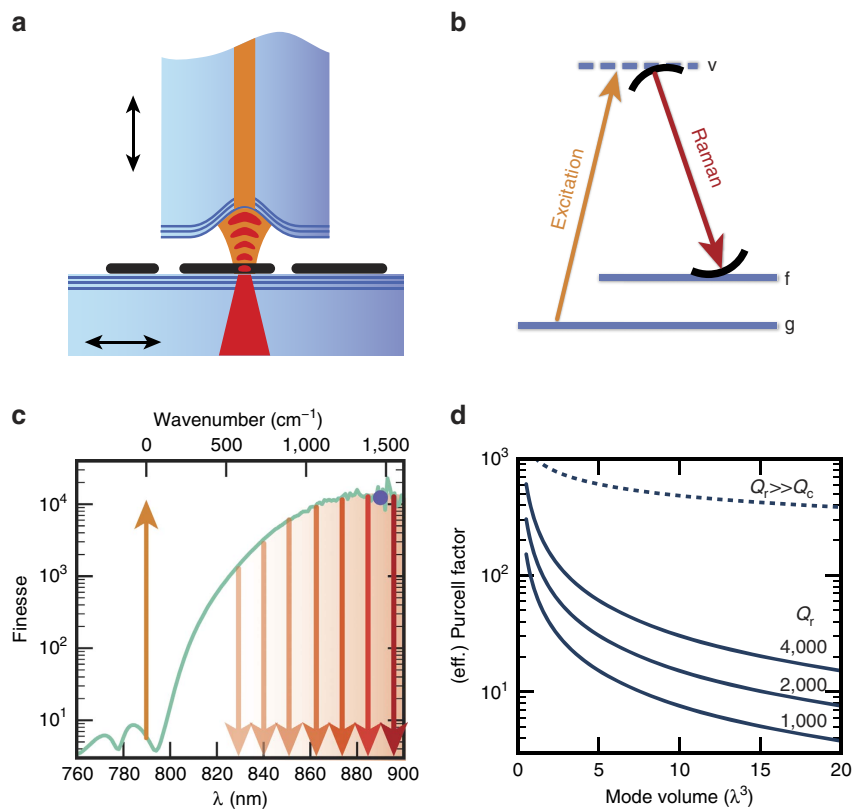

**Figure 1 | Principle of cavity-enhanced Raman spectroscopy. (a)** Schematic of the cavity formed by the end-facet of an optical fibre (top) and a macroscopic mirror (bottom). CNTs (black lines) are deposited on the macroscopic mirror and interact strongly with the light field when inside the cavity. The large mirror can be positioned laterally by a nanopositioner and the cavity length can be tuned by a piezo actuator (see black arrows). **(b)** Schematic level diagram for cavity-enhanced Raman scattering showing a ground state (g), a virtual state (v) and a vibrationally excited state (f). The Raman scattering is stimulated by the resonance with the cavity. **(c)** Calculated cavity finesse from a measurement of the mirror transmission (line) and a direct measurement of the Finesse (blue datapoint) for cavity B (see the 'Methods' section). The upward and downward arrows symbolize excitation and stimulated Raman scattering, respectively. Selective enhancement occurs at a cavity resonance, whose wavelength is set by the cavity length. **(d)** Calculation of the effective Purcell factor as a function of mode volume for different quality factors $Q_r$ of Raman features (solid lines) for $Q_r \ll Q_c$. The dashed line is the ideal Purcell factor, equivalent to the effective Purcell factor for $Q_r \gg Q_c$. It solely depends on the Q-factor of the cavity and is evaluated for cavity B used in the experiments.

**Cavity-enhanced Raman spectroscopy.** Significant additional information can be obtained by recording the Raman spectrum of the sample. To do so, we choose a location where absorption consistent with single CNTs is observed. We excite the cavity with a laser at a wavelength where the mirror coating is transparent (690 nm for cavity A, 786 nm for cavity B) such that light can enter the cavity independently of the mirror separation, see Fig. 1c. The Raman scattered light is red-shifted from the excitation laser to wavelengths where the cavity finesse and thus the Purcell factor is high, such that cavity-enhancement occurs (see Fig. 1b,c). This approach allows to access the range of $\sim 300\text{--}3{,}000 \, \text{cm}^{-1}$, covering the important molecular fingerprint region. To record a broadband Raman spectrum, we stepwise tune the mirror separation to sweep a set of cavity resonances across the desired spectral range where Raman features are expected. The number $k$ of resonances contributing simultaneously to probe the spectrum depends on the effective cavity length $d$ and the considered spectral bandwidth $\Delta \nu$ (in units of Hz) according to $k = \text{ceil}[2d\Delta \nu/c]$. At each spectral position of the cavity resonances we record the Purcell-enhanced scattering rate with a spectrometer and construct a full Raman spectrum by combining the maxima of all spectra. The Raman features of CNTs studied here are the $G^{\pm}$-band related to tangential axial ($+$) and circumferential ($-$) vibrations of neighbouring carbon atoms, the defect-related D-band, and its overtone the $G'$-band[20]. Figure 2a shows a high-resolution measurement of the $G^{\pm}$-band at small mirror separation using a single longitudinal cavity mode. The measurement is combined from 370 individual spectra which are shown in Fig. 2b in a false-colour map as a function of the cavity length. With such a high sampling it is possible to achieve a spectral resolution beyond that of the spectrometer and to resolve also narrow lines, whereby the cavity-enhancement maintains the Raman lineshape. Figure 2c shows a large-bandwidth spectrum where the $G^{\pm}$ and $G'$-bands are visible. Here, resonances with longitudinal mode numbers $q = 34\text{--}38$ contribute due to the larger average mirror separation. For the measurement we combine 30 spectra where we change the cavity length in 10 nm steps and additionally modulate the length with a comparable amplitude at a high frequency. This provides a spectrometer-limited resolution with a reduced number of exposures. The absence of the D-band expected at $1{,}350 \, \text{cm}^{-1}$ indicates high quality of the material. At the same time, the signal is significantly enhanced by the Purcell effect (see below).

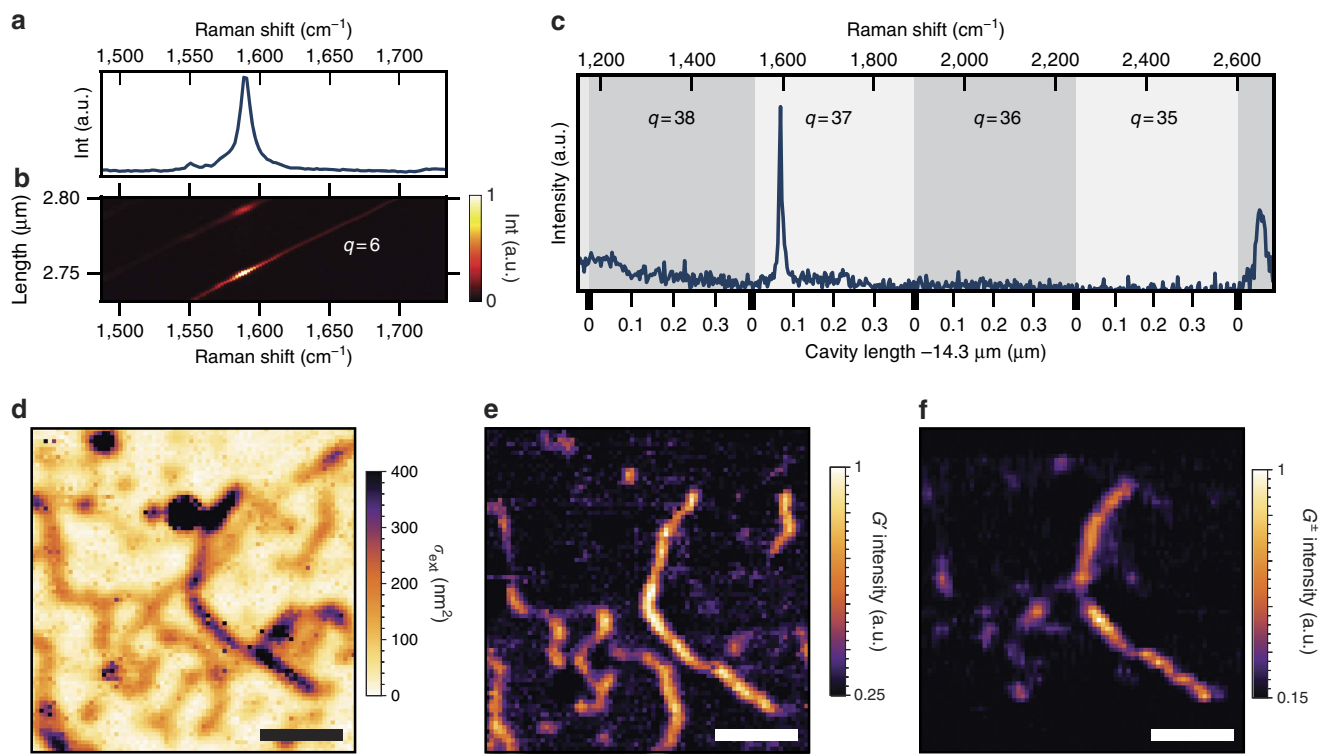

**Figure 2 | Cavity-enhanced absorption and Raman imaging. (a,b)** Acquisition of a Raman spectrum by tuning a cavity resonance across a spectral range around the $G^\pm$ band and combining individual exposures. **(b)** Each horizontal line corresponds to an individual spectrum acquired for a particular cavity length. The strong line shows the tuning of the fundamental mode, the second weaker line corresponds to the first higher transverse mode of the cavity. **(a)** Spectrum obtained by taking the maximum of all exposures for each spectral position. **(c)** Broadband spectrum covering the $G^\pm$ and $G'$ band at 1,590 cm$^{-1}$ and 2,650 cm$^{-1}$ respectively. The differently shaded areas depict the range which individual longitudinal cavity modes cover at a mirror separation of 14.3 μm. **(d)** Scanning-cavity microscopy of the extinction cross-section $\sigma_{ext}$ of isolated CNTs. **(e,f)** Raman hyperspectral imaging of the same area showing selectively the strength of the $G'$ and the $G^\pm$ band, respectively. Scale bars, 20 μm.

**Cavity-enhanced Raman imaging.** We expand this capability to perform hyperspectral imaging. Therefore, we record cavity-enhanced Raman spectra such as the one shown in Fig. 2c at many locations on the mirror. Figure 2e,f show a measurement where the strength of the $G^\pm$ and $G'$-band at 1,590 cm$^{-1}$ and 2,650 cm$^{-1}$ is visualized by selecting spectral bands of about 20 cm$^{-1}$ width, respectively. We observe excellent correspondence of the Raman map and the absorption image. The differences between the $G^\pm$ and $G'$ band partially originate from the spatial variation of sample extinction, which, for this measurement, affects the collected $G^\pm$-band more than the $G'$-band due to the higher finesse at the spectral position of the $G^\pm$ resonance. The spatial resolution of the Raman maps is found to be 2.1 μm, which is higher than for the extinction measurement where it is consistent with the cavity mode waist $w_0 = 3.2$ μm. This derives from the fact that the point spread function for Raman imaging is a product of the excitation and collection point spread functions. Due to a residual low-finesse cavity at the excitation wavelength, both are given by the cavity mode, such that the resolution improves by a factor of $\sim\sqrt{2}$.

Hyperspectral imaging provides a maximum of information about the sample, and can permit one to chart, for example, the diameter distribution of nanotubes (see the subsection 'Optimized hyperspectral imaging' below). The improved signal from the cavity can allow to detect otherwise hidden spatial or spectral structures. The acquisition of a broadband spectrum with the cavity involves several exposures with 100–500 ms integration time at each pixel. Much faster Raman imaging can be realized by scanning-cavity microscopy with a cavity stabilized on resonance with a particular Raman band. The Purcell enhancement of the signal can then provide a net advantage in acquisition time compared with conventional confocal Raman microscopy. We demonstrate this approach in the following using cavity B and a different CNT sample produced in the same way but yielding shorter tubes. We implement a digital optimization algorithm for stabilizing the cavity on resonance with the $G^+$-band of CNTs (see the 'Methods' section). Figure 3a shows an example where we raster-scan the mirror with a cavity length of 3.2 μm and use a dwell time of 400 ms per pixel. Here, the smaller effective radius of curvature of cavity B (see the 'Methods' section) leads to a spatial resolution of 1.5 μm. Individual nanotubes yield a signal of typically $2 \times 10^4$ counts s$^{-1}$ for an excitation power of 14 mW, significantly larger than for a comparable confocal measurement.

**Quantification of Purcell enhancement.** To quantify the cavity-enhancement, we compare the cavity-based measurement of the $G^\pm$ band for the smallest achievable cavity length $d = 2.75$ μm with a corresponding confocal Raman measurement taken with a microscope objective (numerical aperture 0.7) for the same position on the same isolated CNT indicated by a circle in Fig. 3a. We have chosen a location where the extension and orientation of the CNT can be resolved (see the 'Methods' section). We set the polarization of the excitation light to be circular at the sample for both measurements to ensure comparable excitation conditions. Figure 3b shows the signal on the spectrometer for the stabilized cavity and the confocal measurement, normalized to equal excitation intensity at the location of the nanotube. We observe a strongly enhanced Raman signal at the cavity resonance, where the limited spectrometer resolution broadens the cavity line to a non-Lorentzian shape.

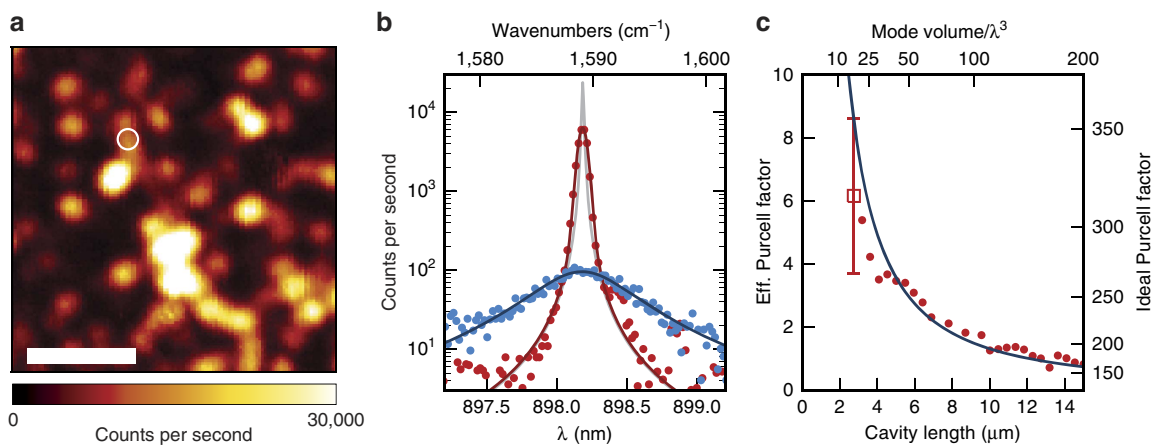

**Figure 3 | Cavity-enhanced Raman imaging and quantification of the Purcell enhancement.** (**a**) Raman map showing the strength of the $G^+$ band with the cavity actively stabilized on resonance with the Raman transition. Scale bar, 10 μm. (**b**) Spectra taken at the location of the circle in **a** with a confocal microscope (blue data points with Lorentzian fit) and the stabilized cavity at a cavity length $d = 2.75$ μm (red data points with Voigt fit) normalized to the same intensity at the nanotube. The grey line shows the intrinsic cavity resonance. (**c**) Scaling of the integrated cavity-enhanced Raman intensity (red data points) as a function of the effective length of the cavity. The red square is evaluated from the measurement shown in **b**, and the solid line is the theoretically predicted ideal and effective Purcell factor. The error bar is the estimated experimental uncertainty (95% confidence interval).

From several such measurements on different CNTs, and taking into account the respective collection efficiencies (see the 'Methods' section), we extract the effective and ideal Purcell enhancement. For the effective Purcell factor we compare the confocal and cavity spectra integrated over the full $G^\pm$ band, yielding $C = 6.2 \pm 2.5$. The error denotes the uncertainty in the determination of the respective excitation intensities and collection efficiencies. A similar statistical uncertainty is obtained when evaluating different nanotubes on the mirror and selecting for high cavity-outcoupling efficiency. The ideal Purcell factor can be obtained from the ratio of the peak spectral densities of the measurements. We extract the peak spectral density of the intrinsic cavity line, which has an independently measured linewidth of $0.15 \, \mathrm{cm}^{-1}$ at the mirror separation used (see the 'Methods' section), by matching its integral with the integrated count rate measured with the spectrometer. This yields an ideal Purcell Factor of $C_0 = 320 \pm 120$. These values are in good agreement with the calculated values $C = 8.5$ and $C_0 = 354$ which were deduced from the cavity mode volume $V = 13\lambda^3$, the cavity quality factor $Q_c = 6.1 \times 10^4$ and the quality factor of the Raman $G^+$ band $Q_r = 1.54 \times 10^3$. The Raman Q-factor was obtained from a fit to the confocally collected Raman line shown in Fig. 3b. The enhancement also follows the expected scaling for Purcell enhancement when varying the cavity length and evaluating the respective integrated count rate, see Fig. 3c.

The enhanced signal is furthermore collected very efficiently. Evaluated at several different locations, we obtain collection efficiencies up to $\eta_c = 60\%$, comparable to the best values achievable with oil-immersion objectives. Overall, the cavity-enhancement and efficient collection provide a net signal gain, which for example can allow a speedup in Raman imaging and enable the detection of weak Raman features, which otherwise might be hidden in the background.

**Optimized hyperspectral imaging.** As a final step, we demonstrate the potential to study the details of spectral features while still benefiting from a net enhancement due to the cavity. We use this for efficient CNT characterization by resolving properties of the $G^\pm$ band, where the line splitting is related to the tube diameter, and the linewidth of the $G^-$ transition allows to separate metallic and semiconducting CNTs[20]. Figure 4a,b show examples for a semiconducting and a metallic nanotube, respectively. For the measurement we weakly stabilize the cavity to an average length where it is resonant with the $G^\pm$ band and modulate the mirror separation with a high-frequency triangular waveform to sweep the cavity resonance across the Raman line. This enables the acquisition of a hyperspectral image covering the $G^\pm$ band. Since we restrict the cavity resonance to a spectral region where a Raman signal is observed, we benefit from the average enhancement, which for the measurement shown still provides more than a factor of three average Purcell enhancement of these lines.

We fit each spectrum with a sum of two Lorentzians and evaluate their separation $\Delta$ and the linewidth of the low-frequency $G^-$ resonance. Figure 4c shows the tube diameter inferred from[20] $d_t = 6.8/\sqrt{\Delta}$, and Fig. 4d shows the $G^-$ linewidth for the same area as shown in Fig. 3a. For most of the nanotubes we obtain a clear indication of their diameter and electronic structure. This technique is particularly robust, since the cavity needs to be stabilized to a mirror separation within a few nanometres only.

## Discussion

We have demonstrated cavity-enhanced Raman spectro-imaging of individual nanoobjects. The accessible spectral range covers the molecular fingerprint region, underlining the general applicability of the technique. Purcell enhancement and the cavity-intrinsic highly efficient collection of signals provide a net gain compared with confocal Raman imaging, and improve the analysis of heterogeneous samples on large areas.

Reducing mode volumes further can improve the Purcell enhancement by about one order of magnitude[32], additional enhancement can be achieved by cavity-enhancement of the excitation laser by up to a factor $10^4$ as shown earlier[9], and collection efficiencies can exceed 90% by slightly optimizing the cavity mirror reflectivities. Overall, enhancement factors beyond $10^6$ can be realized in this way. Excitation of Raman transitions via electronic resonances[20] can additionally increase the signal. Furthermore, the cavity serves as a spectrally selective element without sacrificing signal, such that with suitable calibration, no spectrometer is required. This could enable, cost-effectively, a spectral resolution given by the cavity linewidth, as small as 0.01 cm$^{-1}$. The strong cavity filtering could also allow the acquisition of spectra at very low wave-numbers, for example, by exciting at a shallow angle through the mirror where the dielectric coating becomes transparent. The demonstrated combination of enhancement of Raman signals and ultra-sensitive absorption spectroscopy extends the accessible information compared with other techniques. Finally, high-finesse microcavities should enable low-threshold Raman lasers with a variety of novel materials.

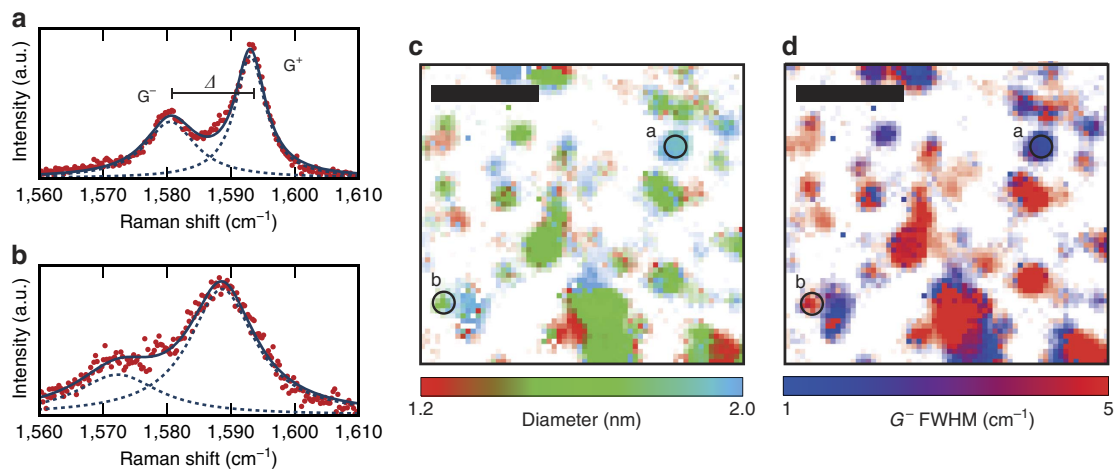

**Figure 4 | Mapping of CNT diameter and electronic structure using cavity-enhanced hyperspectral imaging.** Example of cavity-enhanced G-band spectra for a semiconducting (**a**) and a metallic (**b**) nanotube, taken at the location of the circles shown in **c** and **d**. The solid line is a sum of two Lorentzians fitted to the data and the dashed lines are the underlying individual Lorentzians of the G$^-$ and G$^+$ lines, respectively. (**c**) Spatial map of the evaluated tube diameter from the G$^+$/G$^-$ line splitting measured on the same area as shown in Fig. 3a. An average diameter of 1.5 nm is observed, in good agreement with atomic force microscopy. (**d**) Spatial map of the evaluated G$^-$ linewidth (full width at half maximum, FWHM). Broad lines are expected for metallic nanotubes, while semiconducting nanotubes typically show narrow lines. Scale bars, 10 μm.

Beyond Raman imaging, we believe that our approach could be used to gain optical quantum control over phonons. By harnessing strong coupling between cavity photons and localized excitons in CNTs[29,33,34], Raman transitions could be used to create and detect single phonons[35,36]. This could lead to a hybrid cavity-optomechanical platform[37], where quantum control of mechanical motion becomes accessible on a molecular scale[38].

## Methods

**Optical setup.** The setup combines a scanning-cavity and a confocal microscope, which share a closed-loop nanopositioning stage (attocube ECS 3030) that can translate the sample-carrying mirror between the cavity and the focus of an objective. This enables us to study the same location on the sample with both techniques. For cavity measurements, we couple grating-stabilized diode lasers (690 nm for Raman in cavity A and stabilization in cavity B, 786 nm for Raman in cavity B and cavity characterization in cavity A, 890 nm for cavity characterization in cavity B) into the cavity fibre and collect the cavity transmission with an aspheric lens. The cavity length can be fine tuned across ~2 μm with a shear piezo onto which the fibre is mounted. We detect the transmitted laser light on a sensitive photodiode (Thorlabs APD120) and Raman signals on a spectrometer (Princeton Instruments Acton SP2500 with Andor ikon-M). The confocal setup contains a long working-distance objective with numerical aperture 0.7 (Mitutoyo Plan Apo NIR HR 100x) oriented with an angle of 38° to the surface normal of the mirror, which we account for in the evaluated excitation intensity and collection efficiency.

**Optical microcavity.** We use two different mirror coatings and fibres for experiments covering different wavelength ranges. Cavity A has a coating with a stopband centre at 780 nm with $T_1 = 12$ p.p.m., $T_2 = 60$ p.p.m., $L_1 = 12$, $L_2 = 27$ p.p.m. and a finesse $\mathcal{F} = 2\pi/(T_1 + T_2 + L_1 + L_2) = 56,000$. The fibre of cavity A has an effective radius of curvature $r_c = 190$ μm, leading to $w_0 = 2.7$ μm for a mirror separation of $d = 4.5$ μm as used in the experiment shown in Fig. 2d,e.

There, the quality factor amounts to $Q_c = 2d\mathcal{F}/\lambda = 6.5 \times 10^5$. Cavity B has a coating which is optimized for low background fluorescence and large outcoupling efficiency in the presence of CNT absorption. It is centred at 890 nm with $T_1 = 77$ p.p.m., $T_2 = 335$ p.p.m., $L_1 = 12$ p.p.m., $L_2 = 20$ p.p.m. and $\mathcal{F} = 14,000$. A smaller $r_c = 90$ μm leads to $w_0 = 2.1$ μm for $d = 3.2$ μm. The edges of the fibre B are mechanically polished off to allow for cavity lengths as small as $d = 2.75$ μm, where we obtain a mode volume $V = \pi w_0^2 d/4 = 13\lambda^3$. The finesse of both cavities is measured by probing the cavity transmission with a laser while tuning the mirror separation across more than a free spectral range, and evaluating the resonance linewidth and separation. The effective cavity length $d$, which includes field penetration into the coating, is deduced from the free spectral range observed in broadband cavity-transmission spectra. The mode waist $w_0$ is measured by absorption imaging of point-like objects with the cavity.

**Cavity stabilization.** We use a laser at 690 nm where the mirror has high transmission to provide an error signal for stabilization of cavity B. With the additional transmission signal of the Raman excitation laser at 780 nm, we implement a digital optimization algorithm for locking the cavity on resonance with any Raman transition within the accessible spectral range. Our scheme enables locking also when slip-stick motion of the nanopositioner with significant crosstalk on the cavity length is involved. The mirror separation is stabilized to below 200 pm and therefore maintains the spectral position of the resonance within 0.31 cm$^{-1}$.

**Count rate comparison.** When comparing confocal and cavity-enhanced count rates, we correct for the differences in the respective excitation intensities and collection efficiencies. For the cavity, the outcoupling efficiency through the planar mirror is given by $\eta_c = T_2/(T_1 + T_2 + L_1 + L_2 + 2B)$ and depends on the local extinction of the sample. We use a scanning-cavity extinction measurement covering the area shown in Fig. 3a to calculate $\eta_c$ at the locations used for enhancement determination. The collection efficiency of the objective is calculated by integrating the interfering direct and reflected dipole radiation pattern over the opening angle of the inclined objective, and accounting for the polarization- and angle-dependent complex reflectivity of the Bragg mirror. The obtained value depends on the dipole orientation and amounts to 43% (15%) including objective transmission for orthogonal (parallel) orientation with respect to the projection of the inclined optical axis in the mirror plane. Transmission through further optics is directly measured; confocal and cavity measurements share most of the optical path including filters and coupling to the spectrometer. The excitation intensity at the nanotube in the cavity is calculated from the measured mode waist, the transmitted laser power and the measured and calculated outcoupling mirror transmission. For the confocal measurement, we measure the excitation point spread function and account for the standing wave due to the mirror. Furthermore, when evaluating signals of extended CNTs we account for a spatial overlap factor $\xi$ with the cavity mode, entering the evaluation of $C_0$, $C$ and $B$, which are defined as peak values. The observed average values are, for example, $\langle C \rangle = \xi C$, where for a homogeneous linear CNT larger than the mode diameter $\xi = \sqrt{\pi/8}$, whereas for a point-like sample $\xi = 1$.

**Sample preparation.** Single-wall CNTs were synthesized on $SiO_2$ by chemical vapour deposition following the procedure described in ref. 33. In brief, iron-ruthenium particles spin-coated on $SiO_2$ were used to catalyse the growth of CNTs with methane as carbon feedstock. Before the CVD growth the substrate with catalyst particles was treated by oxygen plasma to yield longer CNTs. Subsequently the substrate was transferred into a standard CVD furnace, heated in an argon–hydrogen (95%/5%) atmosphere to 850 °C and kept in a flow of 1 slm methane and 0.75 slm hydrogen for 10 min before the cool-down in an argon–hydrogen atmosphere. The average diameter of nanotubes obtained with this method was 1.4 nm. The nanotubes were transferred by contact-stamping the $SiO_2$ chip with CVD-grown CNTs onto the mirror substrate. Scanning electron microscopy showed isolated CNTs on the cavity mirror.

**Data availability.** The data that support the findings of this study are available from the corresponding author on request.

## Acknowledgements

We thank Jakob Reichel and Matthias Mader for helpful discussions, Hanno Kaupp for assistance with fabricating the fibres, and Philipp Altpeter and Reinhold Rath for assistance in the clean-room. We gratefully acknowledge financial support by the European Research Council under ERC Grant Agreement no. 336749, the DFG Cluster of Excellence NIM (Nanosystems Initiative Munich), the Center for NanoScience (CeNS) and LMUInnovativ. T.W.H acknowledges funding from the Max-Planck Foundation.

## Author contributions

D.H., A.H. and T.W.H. conceived the experiment. T.H. and D.H. designed and performed the experiments. T.H., D.H and A.H. analysed the data. J.N. and M.S.H. prepared the sample. D.H., T.H. and A.H. wrote the article. All authors discussed the manuscript.

## Additional information

**Competing financial interests:** The authors declare no competing financial interests.

**Reprints and permission** information is available online at http://npg.nature.com/reprintsandpermissions/

**How to cite this article:** Hümmer, T. *et al.* Cavity-enhanced Raman microscopy of individual carbon nanotubes. *Nat. Commun.* **7:**12155 doi: 10.1038/ncomms12155 (2016).